\documentclass[fleqn,usenatbib]{mnras}

\usepackage{newtxtext,newtxmath}

\usepackage[T1]{fontenc}

\DeclareRobustCommand{\VAN}[3]{#2}
\let\VANthebibliography\thebibliography
\def\thebibliography{\DeclareRobustCommand{\VAN}[3]{##3}\VANthebibliography}

\usepackage{graphicx}	
\usepackage{amsmath}	

\usepackage{gensymb}    
\usepackage{comment}    
\usepackage{wasysym}    
\usepackage{ulem}       
\usepackage{placeins}

\usepackage{tikz,xcolor,hyperref}

\definecolor{lime}{HTML}{A6CE39}
\DeclareRobustCommand{\orcidicon}{
	\begin{tikzpicture}
	\draw[lime, fill=lime] (0,0) 
	circle [radius=0.16] 
	node[white] {{\fontfamily{qag}\selectfont \tiny ID}};
	\draw[white, fill=white] (-0.0625,0.095) 
	circle [radius=0.007];
	\end{tikzpicture}
	\hspace{-2mm}
}

\foreach \x in {A, ..., Z}{\expandafter\xdef\csname orcid\x\endcsname{\noexpand\href{https://orcid.org/\csname orcidauthor\x\endcsname}
			{\noexpand\orcidicon}}
}

\title[Formation of terrestrial planets]{Explaining Mercury via a single giant impact is highly unlikely} 

\author[P. Franco et al.]{
P. Franco$^{1,2}$\thanks{E-mail: patrickoliveira@on.br}\orcidP{},
A. Izidoro$^{3,4}$\orcidI{},
O. C. Winter$^{2,5}$\orcidO{},
K. S. Torres$^{2,6}$,
and A. Amarante$^{2,5}$\orcidA{}
\\
$^{1}$National Observatory, Rio de Janeiro, 20921-400, RJ, Brazil\\
$^{2}$Grupo de Dinâmica orbital e Planetologia, São Paulo State University, Guaratinguetá, SP  \\
$^{3}$Department of Earth, Environmental and Planetary Sciences, 6100 MS 126, Rice
University, Houston, TX 77005, USA\\
$^{4}$Department of Physics and Astronomy  6100 MS 550, Rice University, Houston, TX 77005, USA \\
$^{5}$São Paulo State University, Guaratinguetá, 12516-410, SP, Brazil.\\
$^{6}$Centro Federal de Educação Tecnológica de Minas Gerais - CEFET, Curvelo, 35790-000, MG, Brazil\\
}

\date{Accepted XXX. Received YYY; in original form ZZZ}

\pubyear{2022}

\begin{document}
\label{firstpage}
\pagerange{\pageref{firstpage}--\pageref{lastpage}}
\maketitle

\begin{abstract}
The classical scenario of terrestrial planet formation is characterized by a phase of giant impacts among Moon-to-Mars mass planetary embryos. While the classic model and its adaptations have produced adequate analogs of the outer three terrestrial planets, Mercury’s origin remains elusive. Mercury's high-core mass fraction compared to the Earth's is particularly outstanding. Among collisional hypotheses, this feature has been long interpreted as the outcome of an energetic giant impact among two massive protoplanets. Here, we revisit the classical scenario of terrestrial planet formation with focus on the outcome of giant impacts. We have performed a large number of N-body simulations considering different initial distributions of planetary embryos and planetesimals. Our simulations tested the effects of different giant planet configurations, from virtually circular to very eccentric configurations. We compare the giant impacts produced in our simulations with those that are more likely to account for the formation of Mercury and the Moon according to smoothed hydrodynamic simulations. Impact events that could lead to Moon's formation are observed in all our simulations with up to $\sim$20\% of all giant impacts, consistent with the range of the expected Moon-forming event conditions. On the other hand, Mercury-forming events via a single giant impact are extremely rare, accounting for less than $\sim$1\% of all giant impacts. Our results suggest that producing Mercury as a remnant of a single giant impact that strips out the mantle of a differentiated planetary object with Earth-like iron-silicate ratio is challenging and alternative scenarios may be required (e.g. multiple collisions).
\end{abstract}

\begin{keywords}
methods: numerical -- protoplanetary discs -- planets and satellites: terrestrial planets
\end{keywords}



\section{Introduction}

The formation of the terrestrial planets has been an intense topic of research for more than 30 years. The onset of the late stage of accretion of terrestrial planets is characterized by the Sun's natal gas disk dispersal and subsequent planetary growth via a phase of giant impacts among Moon-mass to Mars-mass planetary embryos \citep{W78,W86,CW98,Cham01,Obrien,R06,R09,Morb12,I13,I14,I15,LI13,LI17, Clement18, Clement21}.  In most studies  modeling this stage of planet formation via numerical simulations, the masses and the orbits of Venus and Earth have been broadly matched, while reproducing the masses of Mercury and Mars has been a more challenging problem \citep[e.g.][]{wetherille86,wetherill91,CW98,izidoroetal18}.

Mars-analogues are better reproduced in simulations where the terrestrial planets formed from a disk with a mass-deficit beyound 1~au. This mass deficit either existed from the beginning \citep{Jin08,Hansen09,walsh,I14,I15,I21b,I21a} or was created via different dynamical mechanisms~\citep{nagasawaetal00, Thommesetal08, walsh, Raymondetal2016, Clement18,Brozetal2021}. The small mass of Mars - relative to Earth's and Venus's masses --  is probably a consequence of the lack of mass near its formation location at the time of planetary growth \citep[e.g.][]{wetherill91,Hansen09}. Mercury's small mass also suggests that little material existed inside Venus's current location when Mercury formed \citep[e.g.][]{Hansen09,walsh,Clement_2021,I21b}.
In addition, Mercury is also intriguing from a geological point of view. Its high bulk density of about  5g/cm$^3$  \citep{Urey} suggests that it has a large core-mantle mass fraction ($\sim$70\%; \cite{ES17}), significantly higher than those of other Solar system terrestrial planets, which are expected to be in the range of $\sim$20\% to $\sim$40\% \citep[e.g.][]{LoddersFegley1998,McDonoughetal2021}. Mercury's large core is thought to be the outcome of a violent impact -- of an Earth-mass object with a body just over twice Mercury's current mass -- that removed a significant part of its original mantle \citep{AA04, A10, AR14, SSL,Chau}; or to be linked to preferable local condensation of specific chemical elements as metallic iron~\cite[][]{EbelAlexander2011,Wurmetal2013,Pignataleetal2016,ES17,KrussWurm2018,KrussWurm2020}. This work builds on the impact hypothesis. Table \ref{tab:mercury_models} summarizes the impact configurations existing in the literature that could account for the formation of Mercury via a single impact. These configurations will be later used as reference in our work.

   \begin{table*}
    \centering
        \caption{Summary of the preferred impact configurations proposed in the literature to account for the formation of Mercury. From left-to-right the columns are the reference model,  $M_t$ is the mass of target, $M_p$ is the projectile mass, $V_i$ is the
        impact velocity scaled by the escape velocity ($V_{\mbox{esc}}$), and $\theta_i$ is the impact angle. The bottom row shows the parameters' ranges that will be used in our analysis.}
         \label{tab:mercury_models} 
    \begin{tabular}{lcccc}
    \hline \hline
  Reference	     & $M_{\mbox{t}} (M_\oplus)	$  & $M_{\mbox{p}} \ (M_{\oplus})$  & $V_i \  (V_{\mbox{esc}})$ & $\theta_i$ (deg.) \\ \hline
    \cite{AR14}   & 0.85  & 0.15-0.30	& 1.75-3.5	      & 22-45         \\ 
    \cite{SSL}    & 0.13-0.62  & ~0.13	& 1.5-5	      & 14-74         \\
     \cite{Chau} (Case-2)   & 0.24-0.55 & 0.13-0.2 	& 1-8	      & 10-45         \\ 
     \hline
    Our analysis & 0.1-1 & 0.055-0.3 & 2+ & 10-75 \\
    \hline
       \end{tabular}
        \par
   \end{table*}

The Moon is also thought to be the outcome of a giant impact between the proto-Earth and a Mars-sized proto-planet  \citep{CW76, Benz89, CB91, Ida97, Canup01, CanupAsphaug01, Canup04, Canup08, A14} -- the so called Theia. In order to account for the Moon's estimated low-iron content, it has been proposed that Moon-forming event may have been an oblique low-speed collision. This impact generated a debris disk around the proto-Earth -- the Moon-forming disk -- mostly from material originated from  Theia's mantle, rather than from the iron-rich core, which may have merged with the proto-Earth \citep{CanupAsphaug01, A14}. This scenario is known as the canonical model. Variations in the impact configurations of the Moon-forming event have been also proposed in order to account for other constrains of the Earth-Moon system as, for instance, the system angular momentum \citep{Canup12, CukStewart12, Lock17,Carter19}. Whereas the very impact configuration remains weakly constrained (see Table \ref{tab:lunar_models} for details), there is a wide consensus that the Moon formed via a giant impact between the proto-Earth and another large protoplanetary object.

\begin{table*}
    \centering
        \caption{Summary of the preferred impact configurations of the Moon-forming event in the literature.  From left-to-right the columns are the reference model, $M_t$ is the mass of target, $M_p$ is the projectile mass, $V_i$ is the impact velocity, $V_{\mbox{esc}}$ is the mutual escape velocity and $\theta_i$ is the impact angle. The bottom row shows the parameters' ranges that will be used in our analysis }
         \label{tab:lunar_models} 
    \begin{tabular}{lcccc}
    \hline \hline
  Reference	     & $M_{\mbox{t}} (M_\oplus)	$  & $M_{\mbox{p}} \ (M_{\oplus})$  & $V_i \  (V_{\mbox{esc}})$ & $\theta_i$ (deg.) \\ \hline
    \cite{CanupAsphaug01}   & $\sim$0.9  & $\sim$0.1	& $\sim$1	      & 48-60         \\ 
    \cite{Canup12}   & 0.55-0.65 & 0.4-0.5 	& 1-1.6	      & 20-45         \\ 
    \cite{CukStewart12}    & 0.99  & 0.02-0.1	& 1-3	      & 0-18         \\ 
    \cite{Reu12} & 0.9 & 0.1-0.2  & 1.2-1.25 & 30-35 \\
    \hline
    Our analysis & 0.5-1.1 & 0.015-0.5 & 1-3 & 0-60 \\
    \hline
       \end{tabular}
        \par
\end{table*}

Given the key role of giant impacts on the formation of the Moon and perhaps Mercury, in this work we revisit different scenarios modeling the formation of terrestrial planets in the Solar system aiming to determine how common are collisions that could account for the formation of the Moon and Mercury. We analyze giant impacts in a large suite of N-body numerical simulations. We compare the impact configurations produced in our planet formation simulations with those inferred from smoothed-hydrodynamical simulations -- namely impact geometry, impact velocity, and mass-ratio of the impacting bodies  \citep{CanupAsphaug01,Canup12,CukStewart12,Reu12,AR14, SSL,Chau} - to be more likely to account for the origin of Mercury and the Moon within the giant impact hypotheses (see Table \ref{tab:mercury_models} and \ref{tab:lunar_models}). Our primary goal is to determine how common (or rare) these impact conditions are in numerical simulations testing for different distributions of solids in the terrestrial region and giant planets configurations.

This paper is organized as follows: Section \ref{sec:model} describes our model and approach. Section \ref{sec:resultados} presents the  results from our simulations. Section \ref{sec:final_comments} summarizes our main findings and conclusions.

\section{Models and Numerical Methods}
\label{sec:model}

\subsection{Initial Distribution of Solids}

\label{sec:declive}
 \begin{figure*}
        \centering
        \includegraphics[scale=0.55]{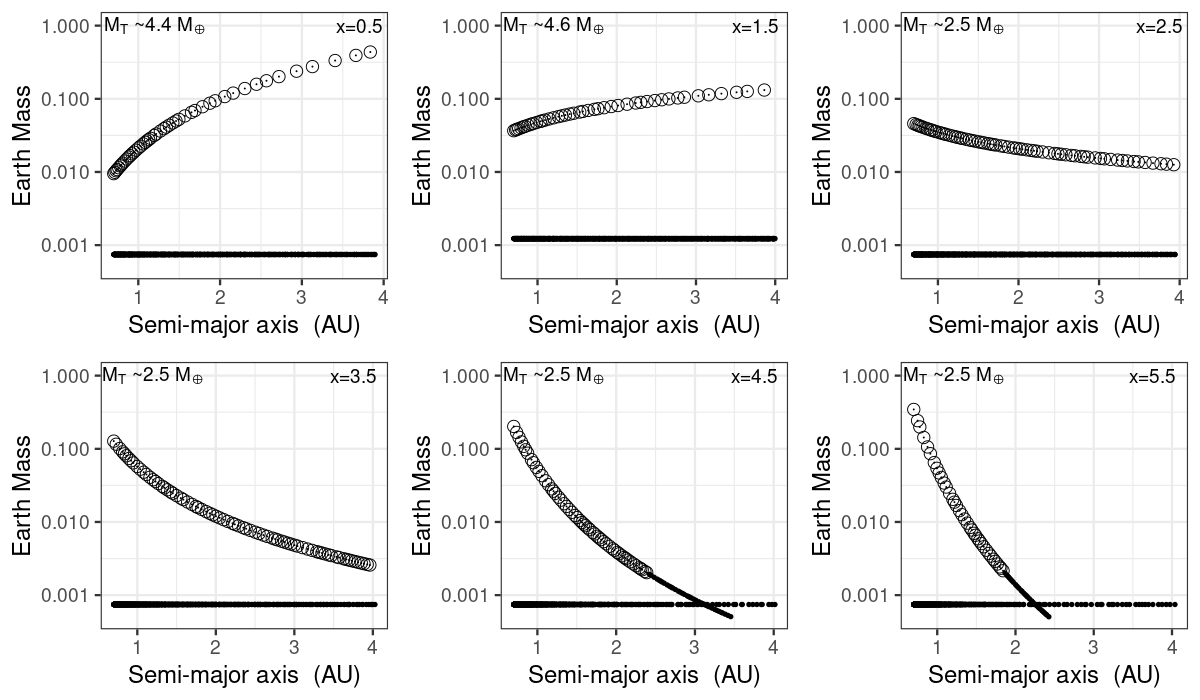}
        \caption {Initial distribution of planetary objects following disks with different surface density profiles.  The horizontal axis shows semi-major axis and the vertical one the initial mass of individual objects. Each panel represents a different surface density slope as indicate by the value of $x$ showed on the top-right of each panel. Planetesimals are represented by dots and planetary embryos by open-circles. The total disk mass (M$_{\text{T}}$), is shown at the top-left corner of each panel.
        }
        \label{fig:slope}
    \end{figure*}
    
We performed 495 N-body  numerical simulations of the late stage of accretion of terrestrial planets. The distribution of solids in the terrestrial region follows  power-law surface density profiles given by $\sum \sim \sum_1r^{-x}$, where  $x$ varies from 0.5 to 5.5 (Figure \ref{fig:slope}). Shallow surface density profiles (lower values of $x$) are consistent with pioneering disk models~\citep{weidenschilling77,hayashi81,R05}. Steep surface density profiles (higher values of $x$) are aligned with planet formation models modeling dust drift due to gas drag in the protoplanetary disk~\cite[e.g.][]{adachietal76,I15,levisonetal15,I21b}. We follow the traditional approach~\cite[e.g][]{R06} of splitting the disk mass between two population of objects, namely protoplanetary embryos and planetesimals  \citep[e.g.][]{KI00}. The protoplanetary disk has a total mass of $\sim$2.5 to 4.5 Earth mass ($M_{\oplus}$), depending on the surface density profile. Following \cite{I15}, disks where $x$ is set to 2.5, 3.5, 4.5 and 5.5 carry $\sim2.5~M_{\oplus}$. Shallow disks ($x=0.5$ and 1.5) carry about $4.5~M_{\oplus}$~\cite[e.g.][]{Cham01}. In all our simulations the disk extends from $\sim$0.7 to $\sim$4 au. Our choice for the disk inner edge at 0.7 au is mostly motivated by previous studies, which show, for instance, that as planetary embryos grow the disk spreads inwards and embryos reach regions as close as $\sim$0.3-0.4 au~\citep{Hansen09,walsh,I15,I21a}. Some models of dust coagulation and planetesimal formation also support an inner edge at  $\sim$0.7 au~\citep[e.g.][]{drazkowskaetal16,I21a}). Previous studies have also considered different scenarios where the disk initially extends  much closer in, down to $\sim$0.3~au~\cite[e.g.][]{Cham01,lykawkaetal19,clementetal21}. This scenario is not explored in this work. 

In our simulations, planetesimals carry initially 30\% to  50\% of the disk total initial mass. Protoplanetary embryos carry the remaining part of the disk mass, and are radially spaced from each other by 3 to 10 mutual hill radii ($\Delta$). Planetary embryos initial masses are set proportional to $M\sim r^{1.5(3-x)}\Delta^{1.5}$, following the runaway and oligarchic growth regimes ~\citep{KI00,KokuboIda2002,R05,R09,I14,I15}. Protoplanetary embryos gravitationally interact with each other, with the Sun, and planetesimals. Planetesimals are not considered to self-interact. 

Planetesimals and protoplanetary embryos have their initial orbital inclinations randomly and uniformly selected  between 10$^{-4}$ and 10$^{-3}$ degrees. Other orbital angles are randomly and uniformly selected between 0\degree ~and 360\degree. Orbital eccentricities are initially set equal to zero. Our simulations were performed using the SyMBA N-body code \citep{Dun98}, and were numerically integrated for 300 Myr, with a time-step of 6 days. When reaching a distance larger than 120 au from the Sun, bodies were considered ejected from the system and removed from the simulation.

\subsection{Giant Planet Orbits}

This paper builds on the so called classical scenario of terrestrial planets formation.  In all our simulations, Jupiter and Saturn are considered fully-formed from the beginning and the Sun's natal disk has already dissipated. We consider a variety of giant planets dynamical configurations, which builds  and expands on previous studies~\citep[e.g.][]{Cham02,Obrien,R09,Obrien14,I14,Clement19b}. Our simulations are performed with
\begin{itemize}
    \item Jupiter and Saturn in 2:1 mean motion resonance (thereafter referred as MMR) and initially with eccentric orbits~\citep{pierensetal14,Clementetal22};
    \item Jupiter and Saturn in 3:2 MMR and initially with circular or eccentric orbits
    \citep[e.g.][]{R09,levison11}.
       \item Jupiter and Saturn initially near their current orbits. \citep[e.g.][]{Cham01,R06}.
\end{itemize}
Table \ref{tab:simu} summarizes the giant planet configurations explored in this work, the total number of simulations of each group, and the slopes of the disk used to build the distribution of protoplanetary embryos and planetesimals.

Our simulations considering cases with Jupiter and Saturn in resonant configurations reflect the view that Jupiter and Saturn probably formed in more compact orbits - due to the tidal interaction with their natal gaseous disk~\cite[e.g.][]{morbidellicrida07}  -- and  reached their current dynamical configuration during an instability phase~\citep{tsiganisetal05,levison11}. The timing of the giant planet dynamical instability is not strongly constrained~\citep{morbidellietal18}, but it probably did not take place later than $\sim$100~Myr after the formation of the Solar system \citep{Nesvornyetal18}. Our simulations with Jupiter and Saturn in resonant orbits reflect scenarios where the dynamical instability is envisioned to have taken place after the terrestrial planets were mostly fully formed (e.g. $\sim$50-150Myr). Simulations considering Jupiter and Saturn near their current orbits, on the other hand, are a good proxy for scenarios where the instability took place early, during the first million of years of the Solar system history~\citep[e.g.][]{Liuetal22}. Note that we do not test the effect of the so-called Solar system dynamical instability itself on the formation of terrestrial planets~\citep[see for instance][]{brasseretal09,Deienno18,Clement18,Clement19a,Nesv21,Sandro21,clementetal21}. However, given the different levels of orbital eccentricity tested in this work our simulations cover, at least at some level, the range of orbital eccentricities that Jupiter and Saturn may have acquired during the instability phase. In the next subsection, we will explain that we test a variety of initial distributions of solids in the terrestrial region. So, our simulations also roughly cover more recent scenarios of terrestrial planet formation proposing that the terrestrial planets accreted from narrow distributions of mass~\citep{Hansen09,drazkowskaetal16,walsh,Raymondizidoro17,I21c}.

\begin{table}
    \centering
        \caption{Setup of the different scenarios covered by our simulations. From left-to-right the columns are the set-name, the number of simulations performed within each set, the initial configuration of the giant planets (Jupiter and Saturn), the giant planets initial eccentricities, and the range of surface density profile slopes for which simulations were performed. We have performed at least 15 simulations for each combination of giant planet configuration, orbital eccentricity and surface density slope.}
         \label{tab:simu} 
    \begin{tabular}{ccccc}
    \hline \hline
     $\text{Simulation}$ & $\text{Number of}$ & $\text{Giant planets}$ & $\text{Giant plaents} $ & $x$ \\
   set &  simulations & $\text{configuration}$ & $\text{Eccentricity}$ &  \\ \hline
    Group 1   & 60    & 2:1 MMR & 0.025 & 2.5-5.5         \\ 
    Group 2   & 60    & 2:1 MMR & 0.050 & 2.5-5.5         \\ 
    Group 3\footnote{}   & 60    & 3:2 MMR & 0 & 2.5-5.5         \\ 
    Group 4   & 60    & 3:2 MMR & 0.025 & 2.5-5.5         \\ 
    Group 5   & 60    & 3:2 MMR  & 0.050 & 2.5-5.5         \\ 
    Group 6   & 75    & Current Orb. & Current Ecc. & 1.5-5.5      \\ 
    Group 7   & 30    & Current Orb. & 0.075 & 0.5-1.5         \\ 
    Group 8   & 90    & Current Orb. & 0.10 & 0.5-5.5          \\ \hline
    \end{tabular}
    \par
    \flushleft{\small{1. Simulation set from  \citet{I15}}}
    \end{table}

\subsection{Treatment of giant impacts}
\label{subsec: regimes}
In order to save CPU-time and make our study doable within a reasonable timeframe, we do not model the effects of collisional fragmentation in our large batch of simulations. Instead, we invoke the reasoning that the treatment of giant impacts as perfect merging events is a fair approximation to model the late stage of accretion of terrestrial planets \citep{KG10, Cham13,Dwyer15,Bonsor15,Carter15,L15,Qui16,Dug19, Clement19a,WL19}. As traditional studies, we solve collisions by assuming mass and linear momentum conservation~\citep{Cham01,R09}. This approach is also supported by the results of simulations modeling the accretion of planets with masses between those of Earth and Neptune  outside the Solar system. These studies also show no statistical difference - in mass and final orbital distribution -- between the result of simulations modeling and neglecting the effects of fragmentation~\citep{poonetal20,estevesetal22}.

Although we do not  solve fragmentation in our simulations, we store the impact conditions of each collision to process the data afterwards. Our goal is to infer which fraction of impacts fall within the favorite range of impact  conditions that could lead to the formation of Moon and Mercury via single giant impacts \citep{Benz88, Benz2007, Canup01, Canup12, CukStewart12, Reu12, A14, AR14, SSL, Lock17, Chau, Carter19}. 
In this work, we do not address the formation of Mercury via multiple impacts \citep[e.g.][]{Chau,clementetal21}. We are unable to explore this scenario because we solve collisions as perfect merging events.

We also compute the fraction of collisions falling within each impact regime as defined in \cite{LS12}. \cite{LS12} performed a suite of smooth-hydrodynamic simulations considering different impact conditions as impact velocity, impact parameter,  target-projectitle mass-ratio, and material composition~\cite[see also][]{AA04,Genda12}. These authors provided fitted solutions that describe the post-collision distributions of mass and velocity of fragments and/or  colliding bodies. Following these authors,  the outcomes of impact can be classified in the following categories:

\begin{itemize}
    \item Merging: $V_i<V_{\mbox{esc}}$;
    \item Graze-and-merge: $b>b_{\mbox{crit}}$ and $V_{\mbox{esc}}<V_i<V_{\mbox{hr}}$;
    \item Hit-and-run: $b>b_{\mbox{crit}}$, $V_{\mbox{hr}}<V_i<V_{\mbox{ero}}$ and $M_{\mbox{lr}} = M_{\mbox{t}}$;
    \item Partial Accretion: $b<b_{\mbox{crit}}$, $V_{\mbox{esc}}<V_i<V_{\mbox{ero}}$, and $M_{\mbox{lr}} > M_{\mbox{t}}$;
    \item Erosion: $V_i>V_{\mbox{ero}}$ and $M_{\mbox{lr}} < M_{\mbox{t}}$,
\end{itemize}
where $V_i$ is the impact velocity and $V_{\mbox{esc}}$ is the mutual escape velocity, $b = \sin(\theta_i)$ is the impact parameter ($\theta_i$ is the impact angle), and $b_{\mbox{crit}} = {R}/{(R+r)}$ is the critical impact parameter, which sets the transition to the grazing impacts. R and r are the target and projectile  physical radius, respectively. $V_{\mbox{hr}}$ and $V_{\mbox{ero}}$ are the hit-and-run and erosion velocities, respectively. $M_{\mbox{lr}}$ is the mass of the largest gravitational remnant and $M_{\mbox{t}}$ is the target mass.

We also follow \citep{LS12} and estimate the mass of the largest remnant (and second largest remnant using the impact-reverse) post impact as:
\begin{align}
    M_{lr} &= \bigg(1-\frac{Q_R}{2Q'^*_{RD}}\bigg)M_{\mbox{tot}},&  &\text{if $0< Q_R/Q'^*_{RD} < 1.8$},   \label{eq:mf1}\\
    M_{lr} &= \frac{0.1}{1.8^\eta}\bigg(\frac{Q_R}{Q'^*_{RD}}\bigg)^\eta~M_{\mbox{tot}},& &\text{if  $Q_R/Q'^*_{RD} \geqslant 1.8$},
    \label{eq:mf2}
\end{align}
where $\eta\sim-1.5$ is a fit-based coefficient and $M_{\mbox{tot}}$ is the sum of the target and projectile masses. $Q_R/Q'^*_{RD}$ is the impact energy scaled by the catastrophic disruption criteria, namely, the specific impact energy required to disperse half of the colliding total mass. We do not compute the mass of the largest fragment for all our collisions but only for selected impacts that may lead to the formation of Mercury or the Moon.

\begin{table*}
 \centering
    \caption{Percentage of collisions in different impact regimes. From left-to-right the columns are disk slope, percentage of perfect merging events, partial accretion, graze-n-merge, hit-n-run, and erosion.}
    
    \begin{tabular}{cccccc}
    \hline \hline
       Disk Slope (x) & P. Merging (\%) &  P. Accretion (\%) &   Graze \& Merge (\%) &   Hit \& Run (\%) &   Erosion (\%) \\ \hline
    0.5 & 5.07 & 55.50 & 13.43 & 23.22 & 2.78 \\
    1.5 & 5.52 & 59.77 & 14.33 & 19.63 & 0.76 \\
    2.5 & 10.66 & 54.72 & 14.49 & 19.55 & 0.57 \\
    3.5 & 12.80 & 57.89 & 12.00 & 16.94 & 0.38 \\
    4.5 & 12.45 & 60.85 & 9.90 & 15.92 & 0.88 \\
    5.5 & 15.03 & 60.83 & 8.31 & 14.71 & 1.12 \\\hline
    \end{tabular}
    
    \label{tab:dep_freqt}
    \end{table*}
   
\section{Results}
\label{sec:resultados}

  \begin{table*}
      \centering
        \caption{Number of collisions in our simulations. From left-to-right the columns show the simulation set, the total number of collisions, the total number of giant impacts, the total number of Mercury-forming events defined as in Table \ref{tab:mercury_models} occurring within 0.7au/Mercury-forming events occurring within 1.5au, the total number of H-\&-R with disrupted projectile (non-necessarily Mercury-forming events), and the total number of Moon-forming events following Table \ref{tab:lunar_models}.}
         \label{tab:col_total} 
        \begin{tabular}{cccccc}
    \hline \hline
    Simulation & Total Collisions & Giant Impacts & Mercury-forming & H\&R-Disr. Projectile  & Moon-forming \\
               &                  &               &     events      &                        &    events                \\ \hline
    Group 1 \\
x=2.5	&10838	&99	    &0/1	&1	&8	\\
x=3.5	&12379	&170	&1/1	&2	&18	\\
x=4.5	&14185	&116	&0  	&0	&16	\\
x=5.5	&15072	&90	    &0	    &0	&22	\\
    \\
    Group 2 \\
x=2.5	&10591	&104	&0	&1	&5	\\
x=3.5	&12395	&167	&0	&1	&13	\\
x=4.5	&14459	&125	&0	&0	&17	\\
x=5.5	&15203	&88 	&0	&0	&16	\\ 
   \\ 
      Group 3 \\
x=2.5	&11704	&104	&1/1	&1	&9	\\
x=3.5	&13500	&164	&0	&1	&15	\\
x=4.5	&12657	&88	&0	&0	&7	\\
x=5.5	&15855	&86	&0	&0	&10	\\
   \\  
    Group 4 \\
x=2.5	&10186	&103	&0	&0	&5	\\
x=3.5	&11642	&161	&0	&0	&16 \\
x=4.5	&13697	&128	&0	&1	&19	\\
x=5.5	&14437	&95	    &0	&0	&23	\\ 
    \\
    Group 5 \\
x=2.5	&7896	&101	&0/1	&4	&14	\\
x=3.5	&9784	&164	&1/2	&3	&20	\\
x=4.5	&11975	&130	&1/2	&3	&18 \\
x=5.5	&12916	&102	&0	    &0	&27	 \\ 
    \\
    Group 6 \\
x=1.5	&6051	&91	    &1/1	&5	&3 \\
x=2.5	&9023	&77	    &0	    &1	&4	\\
x=3.5	&10697	&159	&0	    &1	&14	\\
x=4.5	&13082	&125	&0	    &0	&14	\\
x=5.5	&13918	&99	    &0	    &2	&22	\\ 
    \\
    Group 7 \\
x=0.5	&5395	&119	&1/1	&19	&5\\
x=1.5	&5596	&225	&0/2	&7	&21\\ 
    \\
    Group 8 \\
x=0.5	&5405	&121	&0/3	&6	&4	\\
x=1.5	&5289	&230	&0/4	&8	&24	\\
x=2.5	&6630	&69	    &0  	&0	&2	\\
x=3.5	&8845	&151	&1/1	&1	&19	\\
x=4.5	&11308	&120	&0	    &2	&22	\\
x=5.5	&11186	&90	    &0	    &3	&15	\\ \hline
       \end{tabular}
    \end{table*}

We start this section by presenting the percentage of collisions falling within the different impact categories as defined in the previous section~\citep{SL12}. Next, we quantify the number of Mercury-forming events in our simulations. And finally, we discuss Moon-forming events.
 
 \subsection{Collisional history}

\begin{figure*}
        \begin{center}
            \includegraphics[scale=0.2]{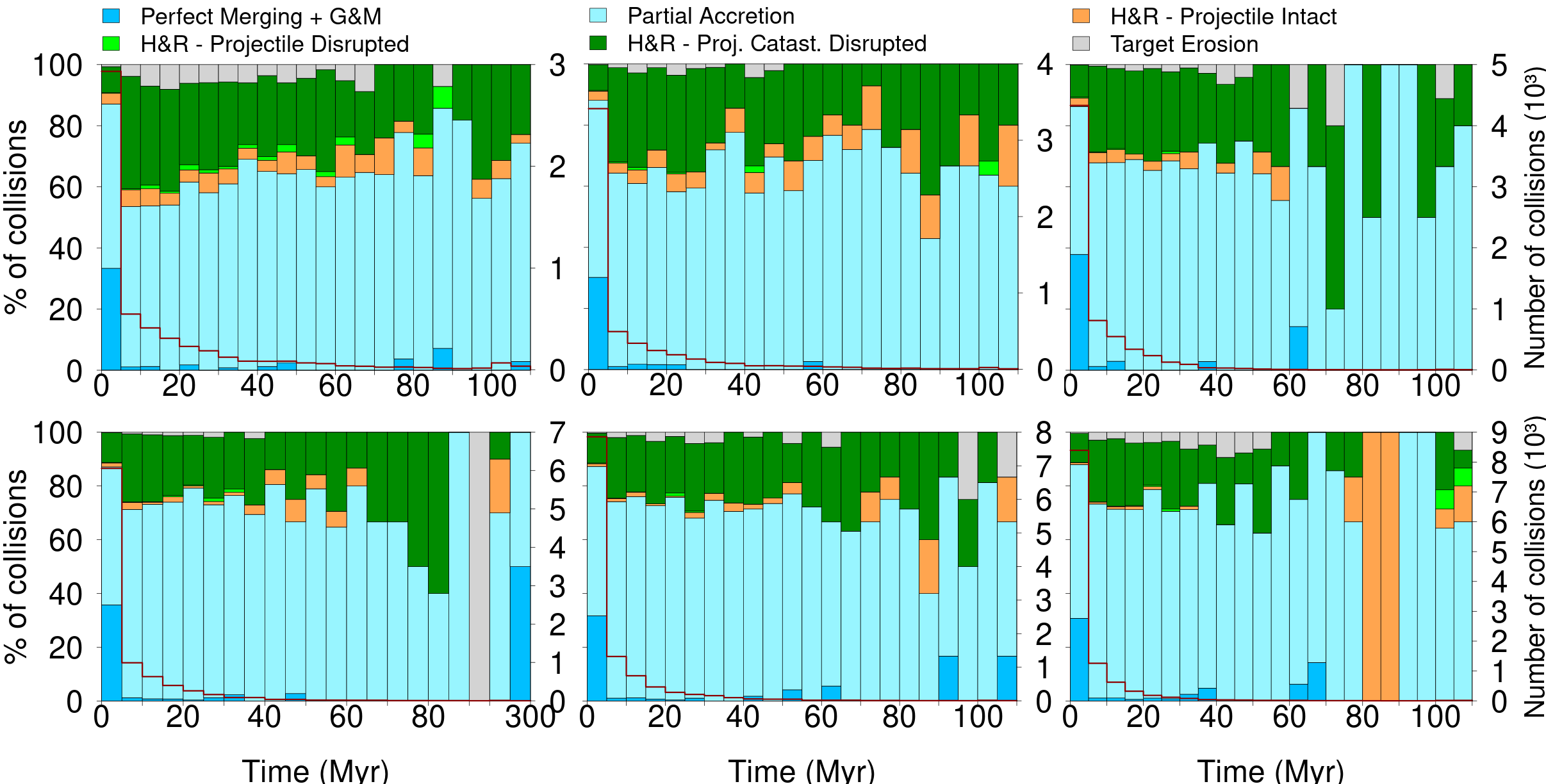}
        \end{center}
        \caption{Statistics on percentage of impacts in different regimes as function of time. The bin-width is set to 5 million up to 100 Myr, and 150 Myr after that. The red histogram line represents the number of collisions over time (right side). Impact data comes from simulations of Group 8. Each panel represents a set of simulations with a different surface density slope, $x$. From left-to-right, top panels represent simulations where $x$ is set to 0.5, 1.5, and 2.5. From left-to-right, bottom panels represent cases where $x$ is set to 3.5, 4.5, and 5.5.} 
        \label{fig:g7_freq}
    \end{figure*}

Table \ref{tab:dep_freqt} shows the percentage of collisions in each impact regime as defined in Section \ref{subsec: regimes}.  One can note that there is no significant difference among the results of all surface density profiles. The overall trends observed in Figure \ref{fig:g7_freq} and Table \ref{tab:dep_freqt} are representative of our simulations. For instance, we note that partial accretion was the predominant type of impact outcome produced in our simulations, accounting for at least 50\% of the impacts. Hit-and-run and Erosion-type collisions corresponds to about 20\% of the simulations in this group. These results are broadly consistent with those obtained by \cite{SL12} for simulations starting with a distribution of Mars-mass planetary embryos (Group 1 of their work). Simulations starting with less massive planetary embryos  (Group 2 and 3 in their work; where planetary embryos had initial masses of 0.005-0.1M$_{\oplus}$) produced a larger number of giant impacts and also more hit-and-run events that disrupted the projectile, compared to our results. This is observed because  a higher mass ratio between colliding objects tend to result in more fragmentation of the smaller body and fewer grazing impacts \citep{SL12}.

 Figure \ref{fig:g7_freq} shows a histogram with the percentage of collisions within each impact regime defined by \cite{LS12}. Each panel represents a set of simulations with different surface density profiles ($x$). All simulations corresponds to Group 8. The histogram-bins width corresponds to time-intervals of 5 Myr. Each histogram-bar is sub-divided and color-coded according to the relative fraction of impacts within each regime at that specific time-interval. As discussed before, most collisions are either in the partial accretion (light-blue) or hit-and-run  regimes (with disruption of the projectile; dark-green). The relative fraction of these two dominant groups does not change drastically during the first 50~Myr of the simulations. After that, large variations mostly come from small number statistics due to the remaining small number of bodies in the simulation.  The red histogram line shows  the number of collisions (see right-hand y-axes) as a function of time. The number of collisions rapidly decreases with time in all scenarios, as expected.

\subsection{Mercury-forming events}
 If Mercury was a differentiated planet in a similar way to other terrestrial planets -- in terms of core-mantle mass fraction --, then it would have to lose considerable mass of the silicate mantle to ensure its current density \citep{ES17}. Different giant impact scenarios have been proposed to account for the formation of Mercury. One of the initial hypothesis proposed that the proto-Mercury was a body of about 4000km of radius (just slightly larger than Mars which has a physical radius of $\sim$3400 km)  that lost $\sim$ 60\% of its mass, mostly from its mantle \citep[][]{Benz88,Benz2007}. Mercury's current radius is $\sim$2500 km and  70\% of its mass is in its metallic core.  \citet{Benz2007} explored this scenario by proposing that  a single head-on giant impact of the proto-Mercury with mass of $\sim$2.25M$_{\mercury}$ and a relatively smaller projectile (1/6 of the target mass \citep{Benz88,Benz2007}) could account for the formation of Mercury. In order to strip a sufficient fraction of the target's mantle, such a collision is required to be sufficiently energetic, and corresponds to nearly 30 times the specific energy suggested for the Moon-forming event in the so called canonical model \citep[][]{Benz2007}. One of the potential issues affecting this scenario is that it may conflict with the abundance of moderately volatile material in  Mercury's current crust~\citep{Sprague95, Kerber09, A10, ES17,Johansen21}  as revealed by the \textit{MESSENGER} mission observations~\citep[][]{Nittleretal2011,Peplowski11}. A very energetic impact would probably raise proto-Mercury's temperature above 1000~K, and this is expected to strongly devolatize the planet, although some volatile material may have been subsequently delivered via leftover planetesimals~\cite[e.g.][]{Morb12,Raymondizidoro17} or accretion of interplanetary dust on long timescales~\cite[e.g.][]{frantsevaetal22}. 
 
 Another argument against the single impact scenario is that producing such an energetic impact requires the projectile to be in an extremely eccentric orbit. However, these sort of events are inferred to be relatively rare in simulations of terrestrial planet formation \citep[e.g.][]{Obrien,Jacksonetal2018, Chau,Clement19b}. Even if possible, subsequent reaccretion of the ejected material is another issue to be avoided \citep[][]{Gladman2009,Spalding2020}. Our large number of simulations and more comprehensive approach -- testing for a variety of different giant planet configurations and mass distributions -- will further constrain the viability of this scenario.

 Table \ref{tab:mercury_models} summarizes the different impact scenarios  proposed in the literature that test in our work, assuming they could lead to the formation of Mercury. We stress again that in this work we focus on the single impact scenario to explain Mercury's high core-mass fraction. The formation of Mercury has been also explored via multiple erosive hit-and-run impacts \citep{AR14, SSL, ES17,Chau} but we are unable to explore this scenario here, as mentioned in the section \ref{subsec: regimes}. The last line of Table \ref{tab:mercury_models}  gives the parameters considered in our analysis, which is based on the different scenarios of the literature. 

In order to quantify how common Mercury-forming collisions are, we scan our 8 sets  of simulations (Group 1 to 8) and we search for impacts that fall within those interval of masses (for both target and projectile bodies), impact velocity, and impact angle defined in Table \ref{tab:mercury_models}. In addition, we impose that the estimated projectile final mass should be comparable to Mercury's mass (within 25\%). In order to estimate the final mass of the projectile we calculate the mass of the second-largest remnant post-impact, following~\cite{LS12}.

Table \ref{tab:col_total} presents the number of collisions occurred in each simulation group and the number of Mercury-forming events (Moon-forming events will be discussed in the next section). The fourth-column of Table \ref{tab:col_total} -- from left-to-right -- shows two entries as ``A''/``B''. ``A'' represents the number of Mercury-forming events -- defined as in Table \ref{tab:mercury_models} --  taking place at heliocentric distances smaller than 0.7~au and at ``B'' the respective number taking places at distances smaller than 1.5~au. Overall, the number of Mercury-forming events inside 1.5~au in our simulations is extremely low, with no more than 4 impacts in a given  simulation set. By restricting our sample to only giant impacts taking place inside 0.7~au this number is never larger than 1. Even in our more extreme scenarios, which assume giant planets on more eccentric orbits had no better-success in producing Mercury-forming events compared to scenarios with less eccentric giant planets. 

Summarizing, our simulations suggest that it is unlikely that Mercury is a remnant of a stripped projectile that experienced a hit-and-run impact event with an Earth-mass object. Sufficiently energetic impacts following this specific design were not observed in any of our simulations. If Mercury is indeed the outcome of a hit-and-run impact, it is more likely to be a consequence of a collision where the target is just a few times more massive than the impactor (M$_{\mbox{tg}}\sim$M$_{\text{proto}\mercury}$), as proposed by \cite{SSL} and \citet{Chau} (Case-2). Mercury-forming events listed in Table \ref{tab:col_total} do not involve objects more massive than a few Mars-mass. Mercury-forming events ocurring inside 0.7~au correspond to  less than $\sim1\%$ of all giant impacts in our simulations.  This fraction does not increase significantly even if we generously consider all collisions taking place inside 1.5~au.  Therefore, our results strongly suggest that forming Mercury via a single energetic giant impact is unlikely.

Multiple hit-and-run erosive impacts remains as a potential alternative solution to this problem, although it may not be able to explain Mercury's volatile content and/or to avoid subsequent re-accretion of the ejected material \citep{Cham13,Chau}. Future studies should further explore this scenario.

\subsection{Moon-forming events}

The canonical model for the formation of the Moon suggests that the Earth-Moon system would have formed from a slight energetic collision ($V_i \sim V_\text{esc}$), between bodies with a 9:1 mass ratio \citep{Canup04}. The giant-impact hypothesis for the Earth-Moon system is in agreement with several constraints~\citep{A14}.

As in the case of Mercury, the origin of the Earth-Moon system has been extensively studied via smoothed hydrodynamic simulations~\citep{Canup12, CukStewart12, Reu12,Lock17,Carter19} and N-body terrestrial planet formation models \citep[e.g.][]{Kaib_Cowan2015, Quarles_Lissauer2015, Jacobsonetal2017}. We summarized in Table \ref{tab:lunar_models} the preferable impact conditions that previous studies have found to be consistent with the formation of the Earth-Moon system. In our analysis, we adopted a generous range of possibilities by combining the range of parameters of the different models presented in Table \ref{tab:lunar_models}. With our generous definition of Moon-forming event in Table \ref{tab:lunar_models} we scan our large sample of simulations and count for the number of collisions matching those parameters. We report our results in Table \ref{tab:col_total}.

The last column of Table \ref{tab:col_total} shows the number of Moon-forming events following the definition of Table \ref{tab:lunar_models}. Our simulations produce impact conditions within the range expected for the Moon-forming event in all groups of simulations and disks, including the eccentric giant planets disks that overall have more energetic collisions. We observed that the number of lunar collisions tends to be greater in the steeper disks -- 10-30\% of all giant impacts (while 5-10\% of giant impacts in shallow disks). We find no obvious trend between the results of our different scenarios affecting the number of Moon-forming events. 

\subsubsection{The timing of the last giant impact on Earth}

For completeness, we now analyze the correlation between the timing of the last giant impact on Earth-analogues and the disk initial surface density slope (x). The Moon-forming event must be the last giant impact experienced by Earth, because both Earth and Moon have an almost identical isotopic composition \citep{jacob14}. The timing of the Moon-forming event is constrained to have taken place between 30 and 150~Myr after the formation of the Solar system~\citep{Yinetal2002,Jacobsen2005,Toubouletal2007,Allegreetal2008,Tayloretal2009,jacob14,mauriceetal20}. We define the last giant impact on Earth-analogues following the same criteria adopted in  \citet{I15}. Earth-analogues have semi-major axes between 0.7~au and 1.2~au
and masses larger than  0.6 M$_\oplus$. The last giant impact on Earth-analogues are those where the impactor mass is larger than 0.026 M$_\oplus$ (or M$_{\mbox{tg}}$).

\begin{figure*}
    \centering
    \includegraphics[scale=0.7]{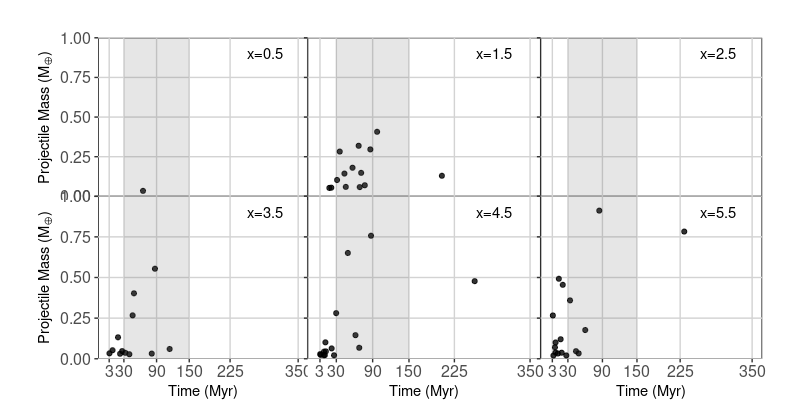}
    \caption{Timing of the last giant impact on Earth-analogues  in simulations of Group 8. The disk slope is shown at the top-right of each panel. The grey area represents the expected range (30–150 Myr) for the last giant impact on Earth based on cosmochemical studies.}
    \label{fig:last_impact_slope}
\end{figure*}

Figure \ref{fig:last_impact_slope} shows the timing of the last giant impact on Earth-analogues in our simulations of Group 8. There is a clear trend in our results. Simulations with shallower surface density profiles tend to have the last giant impact later than simulations with steeper disks. About 80\% of the last giant impacts on Earth analogues in disks with $x=0.5-3.5$ took place between 30 and 150~Myr. More steep disks had about 60\% of the collisions taking place as early as $\sim$10-30 Myr. Yet, $\sim$40\% of the last giant impacts in the steepest disk scenarios (e.g. $x$=5.5) are consistent with the timing of the Moon-forming event as constrained by cosmochemical studies (see Figure \ref{fig:last_impact_slope}). Disks with more steep surface density profiles have most of its mass in the innermost regions. Consequently, terrestrial planets tend to form relatively faster in these disks than in shallow ones \citep{R05,I15}.

\section{Final Comments}
\label{sec:final_comments}
 In this paper, we used a large number of numerical simulations modeling the late stage of accretion of terrestrial planets to study the outcome of impacts, with focus on the formation of Mercury and the Moon. Our simulations are conducted following the classical scenario of terrestrial planet formation. We test the effects of different giant planet configurations and initial distributions of planetary embryos and planetesimals, representing distributions of solids with steep and shallow surface density profiles. In our simulations, Jupiter and Saturn are assumed fully formed and with different levels of orbital eccentricities, varying from almost circular (e$\sim$0) to very eccentric orbits (e$\sim$0.1). Our simulations cover a large range of the parameter space of the classical scenario.
 
 In this work, we do not model the effects of collisional fragmentation~\cite[see for instance][]{KG10,Cham13,poonetal20,estevesetal22}. In order to save CPU-time, we model collisions as perfectly inelastic events, which conserve mass and angular momentum. Yet, we store the impact data of our simulations (target and projectile masses, impact angle, impact position, and impact velocity) in order to  post-process our results and infer the frequency of collisions that could lead to the formation of Mercury and the Moon. Our analysis of impacts follow the empirical fits from \cite{LS12}. We also compare the impacts of our simulations with the preferable impact conditions to account for the formation of the Moon and Mercury according to the results of smoothed-hydrodynamical simulations  \citep{Benz88, Benz2007, Canup01, Canup12, CukStewart12, Reu12, A14, AR14, SSL, Lock17, Chau, Carter19}.
 
Among our results we highlight the following:
\begin{enumerate}
    \item Our results shows that the formation of Mercury via a giant impact that strips out a large fraction of the mantle of a differentiated planetary embryo is a low probability event, at 1\% level or lower.
    \item Venus and Earth-mass planets were produced in all our simulations with different initial distribution of solids and giant planet configurations. Our simulations also show that steeper surface density profiles tend to form planets faster than  shallow disks. These results are in agreement with those of \citet{R05} and \citet{I15}. Up to 30\% of all giant impacts produced in our simulations are potentially consistent with that expected for Moon-forming event.
    
\item Our simulations with steep surface density profiles produced Mars-mass planets around 1.5~au regardless of the giant planet configuration. Simulations with shallow surface density profiles only formed Mars-mass planets around 1.5~au in cases where the giant planets had initially eccentric orbits~\citep{R09}

\item Disks with Jupiter and Saturn in resonant orbits produced lower AMD values than the actual inner Solar system AMD values, whereas disks with very eccentric giant planets produced high AMD values. Also, there was a clear trend emerging from disk with different slopes. Shallow disks produced systems with higher AMD values than steep disks. This difference is due to the initial distribution of bodies in the disk which enhances the effects of dynamical friction in the steep case scenario~\citep{Obrien,R06,I15};
\end{enumerate}

We conclude by stating that our work do not produce good solar system analogues in any of our simulations. Our simulations suffer from many of the fundamental problems affecting the classical scenario of terrestrial planet formation~\citep{R09,I15}. Our work also does not entirely disprove the hypothesis that Mercury's high core-mantle mass fraction  is  the outcome of a single energetic giant impact. However, they clearly show that the preferable impact conditions from smoothed hydrodynamic simulations to account for the formation of Mercury are extremely rare events. This is consistent with the results of previous studies of the late stage of accretion of terrestrial planets~\citep[e.g][]{Clement19b,Clementetal22}. Even in simulations where the final planetary systems are far more dynamically excited than the current Solar system (which would be a thermometer for more energetic impacts during accretion) the success rate is dramatically low. On the other hand, the conditions required to reproduce the Earth-Moon system were observed in  all our simulations and it accounts for a significant fraction of impact events. It is legitimate to wonder if the relatively lower fraction of Mercury-forming events in our simulations compared to Moon-forming events is a simple consequence of more restrictive impact scenarios for the origin of Mercury than for the well-studied Earth-Moon system. We do not believe that this is the case. Instead, we believe that the required impact velocities to produce Mercury via single impact are just too high, and rarely observed in simulations of the late stage of accretion of terrestrial planets.

\section*{Acknowledgements}

We are very grateful to the referee, Matt Clement, for his very constructive review that helped us to improve the manuscript. A.I. is also grateful to Sean Raymond for stimulating conversations and support at the time when these simulations were performed. This work was supported by Fapesp (proc. 2016/24561-0, 2016/12686-2, and 2016/19556-7 ) and CNPq (proc. 305210/2018-1).  A.~I. acknowledges support from The Welch Foundation grant No. C-2035-20200401, NASA grant 80NSSC18K0828 (to Rajdeep Dasgupta),   and the Brazilian Federal Agency for Support and Evaluation of Graduate Education (CAPES), in the scope of the Program CAPES-PrInt, process number 88887.310463/2018-00, International Cooperation Project number 3266.


\section*{Data Availability}
The data underlying this article were provided by André Izidoro under by permission. Data will be shared on request to the corresponding author with permission of André Izidoro.


\bibliographystyle{mnras}
\bibliography{references} 




\newpage
\appendix
\section{Example of final planetary systems}

\begin{figure*}
\begin{center}
        \includegraphics[scale=0.5]{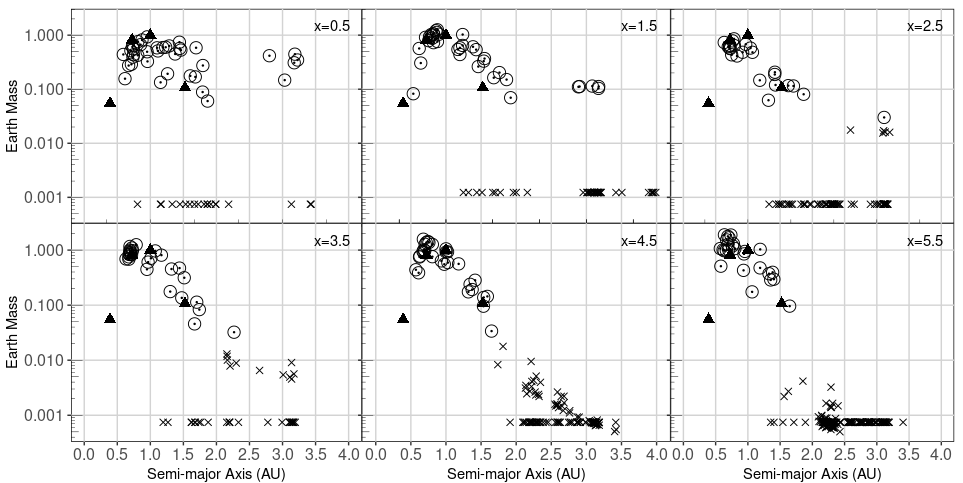}
\end{center}
    \caption{Final mass distribution of the remnant bodies at the end of integration in slope model simulations (Group 8). The surface density slope value, $x$, is indicated on the upper-right corner of each panel. Bodies with masses larger than 0.3M$_\oplus$ are represented by open circles and the crosses represent the smaller bodies. The masses of the terrestrial planets of the Solar system are represented by the solid triangles.}
    \label{fig:sm_s}
\end{figure*}

Figure \ref{fig:sm_s} shows the distribution of semi-major axes and masses of planets produced in simulations of Group 8. By comparing all these panels, one can clearly see that steeper  disks  (higher values of $x$) tend to produce less massive planets further than 1-1.5 au. This result is consistent with previous simulations showing  that the low mass of Mars probably reflects a mass deficit beyond 1-1.5~au at the time of Mars' formation~\citep{Hansen09,walsh,I15,levisonetal15}. This simulation group also produces Earth and Venus-mass planets near 0.7 and 1~au. However, in all these sixty simulations, only one Mercury-mass planet was produced inside 0.5~au (middle top panel, for x=1.5). This difficulty in producing low-mass planets inside 0.5-0.7~au is, at least at some level, an artifact created by the initial conditions of our simulations. As one can see in Figure \ref{fig:sm_s}, some of our simulations -- depending on the value of $x$ --  start with protoplanetary embryos inside 1~au up to a few times more massive than Mercury. Naturally, these simulations can not form or struggle to form Mercury-mass planets because we model collisions as perfect merging events. We will later post-process the impact conditions of our simulations in order to quantify the fraction of impact events that could have accounted for the formation of Mercury via a single erosive giant impact, if fragmentation was accounted for. We will also show that most collisions fall into the partial accretion regime, which is consistent with previous studies~\cite[e.g.][]{SL12}.
 
The results of Figure \ref{fig:sm_s} are qualitatively representative of our full set of simulations. The steepest disks tend to produce overall more massive planets around Venus and Earth regions than shallow disks~\cite[see also][]{I15}. Mercury-mass planets ($M_{\mercury}\pm50\%$) were also produced inside 0.5~au  in Group 1 ($x=$ 3.5 and 4.5), Groups 2 and 3 with $x=$ 5.5 \citep{I15}, Group 4 ($x=$ 5.5), and Group 5 ($x=$ 2.5), but in most cases only 1 final planet fall in this category for each set. Mars-mass planets  ($M_{\mars}\pm50\%$)  were also produced in Group 1 ($x=$ 3.5 and 5.5), Groups 2 and 3 ($x=$ 5.5), Group 4 with $x=$ 4.5 and 5.5 \citep{I15}, Group 5 ($x=$ 5.5), Group 6 ($x=$ 1.5, 3.5, and 4.5), and Group 7 ($x=$ 5.5). The low-mass of Mars either requires a steep surface density profile or giant planets on eccentric orbits to deplete the Mars-region efficiently~\citep{R09,I15,Clement19a}.
 
We now use the angular momentum deficit (AMD) and radial-mass concentration metrics (RMC) to assess quantitatively the orbital configuration of our final planetary systems. The AMD is the difference between the angular momentum of the planet's orbit  and an orbit with the same semi-major axis, but with orbital eccentricity ($e$) and inclination ($i$) equal to zero \citep{Cham01}, and it is defined as
\begin{equation}
    \text{AMD} = \frac{\sum\Big[m_j\sqrt{a_j}\Big(1-\sqrt{1-e_j^2}\cos i_j\Big)\Big]}{\sum m_j\sqrt{a_j}},
\end{equation}
where $m_j$, $a_j$, $e_j$, and $i_j$ are the mass, semi-major axis, orbital eccentricity and inclination of j-th planet, respectively. 

The RMC (the radial mass concentration statistics) measures the degree that a mass is concentrated within one region of a planetary system \citep{Cham01}. The value of RMC varies as a function of $a$ \citep{Cham98, Cham01, R09, I14, I15} and is defined as
\begin{equation}
    \text{RMC} = \max\bigg\{\frac{\sum m_j}{\sum m_j[\log_{10}(a/a_j)]^2}\bigg\}.
\end{equation}
Following \citet{I14,I15}, our calculation of AMD and RMC take into acccount only planets more massive than 0.03$M_\oplus$ and with semi-major axis between 0.3 and 2~au.

Table \ref{tab:oc_s} shows the mean number of planets, mean AMD, and mean RMC of the planetary systems produced in our simulations. In some simulation groups, the mean RMC value was comparable or even higher than that of the terrestrial planets in the Solar system. That occurs especially in simulations with Jupiter and Saturn initially on (more) eccentric orbits. In these specific simulations (e.g. Group 8 where the giant planets are initially very eccentric), fewer planets tend to form between 0.3 and 2 au, consequently the RMC may be higher. Whereas the RMC is higher in these cases, the mean AMD of planetary systems in these simulation are also dramatically high, except for some particular cases where the giant planets start with orbital eccentricities lower or comparable to the current ones  (Group 1, $x=$ 5.5 and Group 5, $x=$ 5.5). Simulations with giant planets starting with orbits more eccentric than the current ones also tend to produce dynamically excited planetary systems. This suggests that although the orbits of Jupiter and Saturn may have been more eccentric in the past~\citep{pierensetal14,Clement18}, such a phase is unlikely to have lasted long (e.g $>$ 10~Myr). This is particularly true if the asteroid belt was born with a low mass from the beginning~\citep{I21b,I21a}, as in our disk with steep surface density profiles. In simulations with shallow surface density profiles, the effect of the giant planet eccentricities are attenuated because the scattering of objects from the asteroid belt -- which contains more mass --  helps to damp the eccentricity of the giant planets, as a consequence of angular momentum conservation~\citep[e.g.][]{R09}. 

Although our simulations do not produce good Solar system analogues in terms of dynamical excitation, we consider our simulations qualitatively adequate to probe the feasibility of producing Mercury via giant impacts given the large range of dynamical excitation of our systems.

\newpage
\begin{table*}
        \caption{From left-to-right  columns show the simulation set, mean number of planets,  mean AMD, and mean RMC. The Solar system is shown at the bottom for reference. The values between parentheses show the interval range over which the mean was calculated.}
         \label{tab:oc_s} 
 \resizebox{0.66\textwidth}{!}{
\begin{centering}
 \begin{tabular}{cccc}
    \hline \hline
    Simulation & Mean N & Mean AMD $(\times10^{-3})$ & Mean RMC\\ \hline
    Group 1 \\
    2.5   & 3.00 (2-4)   & 2.3 (0.4-7.7) & 56.85 (38.66-85.74)         \\ 
    3.5   & 3.27 (2-4)   & 2.2 (0.2-11.2) & 51.88 (29.98-75.18)         \\
    4.5   & 3.67 (2-5)   & 1.8 (0.1-8.1) & 58.94 (35.39-81.99)         \\ 
    5.5   & 3.60 (2-5)   & 1.7 (0.1-9.2) & 80.98 (56.80-245.29)         \\ 
    \\
    Group 2 \\
    2.5   & 3.13 (2-5)   & 4.5 (0.5-31.3) & 53.13 (30.59-82.51)         \\ 
    3.5   & 3.73 (2-5)   & 3.2 (1.1-10.8) & 55.84 (37.37-103.17)        \\
    4.5   & 3.67 (2-5)   & 4.1 (1.0-12.6) & 58.27 (33.68-96.03)         \\ 
    5.5   & 4.20 (3-5)   & 1.8 (0.5-3.5) & 69.27 (53.70-89.16)         \\ \\
    
    Group 3 \\
    2.5 & 2.93 (2–4) & 4.7 (0.3–24.6) & 60.25 (33.33–118.19) \\
    3.5 & 3.53 (2–5) & 2.0 (0.5–5.8) & 50.40 (41.63–62.33) \\
    4.5 & 4.33 (2–6) & 1.2 (0.1–8.4) & 52.23 (42.20–60.47) \\
    5.5 & 4.73 (4–6) & 0.4 (0.1–1.6) & 57.15 (45.87–65.13) \\ \\
    
    Group 4 \\
    2.5   & 3.07 (2-4)   & 2.6 (0.5-9.4) & 51.59 (34.58-82.63)         \\ 
    3.5   & 3.07 (2-4)   & 2.3 (0.1-5.7) & 56.05 (37.25-84.00)         \\
    4.5   & 3.20 (2-4)   & 2.2 (0.3-6.0) & 67.30 (48.05-120.35)         \\ 
    5.5   & 3.46 (2-5)   & 1.8 (0.1-10.8) & 72.40 (58.97-87.24)         \\ \\
    
    Group 5 \\
    2.5   & 2.92 (2-4)   & 5.0   (0.4-18.2) & 100.36 (48.58-178.91)         \\ 
    3.5   & 2.67 (2-3)   & 3.2   (0.4-13.8) & 142.68 (48.36-309.49)         \\
    4.5   & 2.86 (2-4)   & 3.6   (0.1-19.1) & 102.34 (58.77-192.62)         \\ 
    5.5   & 2.64 (2-4)   & 1.7   (0.3-7.3) & 126.04 (85.61-268.80)         \\ \\
    
    Group 6 \\
    1.5   & 2.57 (2-4)   & 11.8   (2.1-32.8) & 77.83 (28.26-313.41)          \\
    2.5   & 3.40 (2-5)   & 4.1   (0.2-28.8) & 75.25 (52.52-123.19)         \\ 
    3.5   & 3.13 (2-4)   & 4.0   (1.2-11.6) & 60.46 (43.70-84.18)         \\
    4.5   & 3.00 (2-5)   & 3.5   (0.2-18.3) & 70.66 (33.48-124.63)         \\ 
    5.5   & 2.79 (2-4)   & 2.6   (0.2-5.3) & 77.56 (47.45-108.20)         \\ \\
    
    Group 7 \\
    0.5   & 2.69 (2-4)   & 15.9   (0.9-53.6) & 91.57 (41.26-394.27)         \\ 
    1.5   & 2.42 (2-3)   & 8.8   (0.2-23.9) & 133.15 (48.13-544.09)         \\ \\
    
    Group 8 \\
    0.5   & 2.43 (2-3)   & 19.0   (0.9-128.7) & 73.85 (19.53-147.06)         \\
    1.5   & 2.82 (2-4)   & 8.8   (1.6-21.8) & 92.28 (48.00-307.58)          \\
    2.5   & 3.29 (2-3)   & 35.2   (4.4-79.1) & 103.75 (55.04-214.53)         \\ 
    3.5   & 3.00 (2-3)   & 10.5   (3.6-27.3) & 74.47 (53.14-104.35)         \\
    4.5   & 3.33 (2-4)   & 10.2   (4.3-35.1) & 96.33 (61.95-160.21)         \\ 
    5.5   & 3.29 (2-4)   & 19.1   (0.7-62.9) & 86.61 (39.32-163.63)         \\ \hline
    
    SS & 4    & 1.8 & 89.9 \\  \hline
    \end{tabular}
    \end{centering}
    }
    \end{table*}


\bsp	
\label{lastpage}
\end{document}